\documentclass{optica-article}

\journal{opticajournal} % for journals or Optica Open

\articletype{Research Article}

\usepackage{lineno}
\usepackage{ulem}
\renewcommand{\sout}[1]{}
\newcommand{\rev}[1]{{\color{black} #1}}

\begin{document}

%\title{Understanding the Complex Time Delay of Chirped Pulses}
\title{Chirped Pulse Analysis and Control in Non-Hermitian Scattering Systems using Complex Time Delay}
% Alternative titles:
% Taking Control of Complex Time Delay Through Chirped Pulses
% Controlling Complex Time Delay Through Chirped Pulse Shaping
% Chirped Pulse Propagation and Control in Reverberant non-Hermitian Settings Enabled Through Complex Time Delay

\author{Isabella L. Giovannelli,\authormark{1,*} Steven M. Anlage,\authormark{1} and Thomas M. Antonsen\authormark{2}}

\address{\authormark{1}Quantum Materials Center, Department of Physics University of Maryland, College Park, Maryland 20742, USA\\
\authormark{2}Department of Electrical and Computer Engineering University of Maryland, College Park, Maryland, 20742, USA
}

\email{\authormark{*}igiovann@umd.edu} 

%%%%%%%%%%%%%%%%%%%%%% Abstract %%%%%%%%%%%%%%%%%%%%%%%%%

\begin{abstract*} 
 We theoretically and experimentally establish a connection between linearly chirped pulse propagation properties and \rev{the complex generalization of Wigner-Smith time delay} \sout{complex time delay} for both transmitted and reflected pulses in linear and dispersive reverberant \rev{non-Hermitian} scattering systems. We demonstrate that the time shift of the chirped pulse depends on both the real and imaginary parts of the complex time delay of the scattering system. We also show that the chirped pulse experiences a center frequency shift that is directly proportional to the imaginary component of complex time delay, similar to that found in Ref.\,\cite{Giovannelli2025}. Using these insights, we then demonstrate how complex time delay can be harnessed to systematically tune the propagation properties of a chirped pulse such that a near-zero time shift can be achieved for a wide range of pulse center frequencies in a resonant scattering system. \rev{Overall, this work broadens the utility and establishes the physical significance of complex time delays in non-Hermitian settings.}
\end{abstract*}

%%%%%%%%%%%%%%%%%%%%%%%%%%  Body  %%%%%%%%%%%%%%%%%%%%%%%%%%

%%%%%%%%%%%%%%%%%%%%%%%%%% Intro %%%%%%%%%%%%%%%%%%%%%%%%%%%
\section{Introduction}
Chirped pulses are pulses whose instantaneous frequency varies throughout the pulse. They are the basis for multiple modern technologies and are present in various physical phenomena. A nonexhaustive list includes chirped pulse amplification \cite{Strickland1985}; chirped pulse interferometry \cite{Kaltenbaek2008,Berube2025} and spectroscopy \cite{ParkCPFTMS2016,Markmann2023}; lidar, radar, and sonar \cite{Allen2001,Adany2009,Saperstein2007,Piracha2011,Darlington1954,Klauder1960,Lihachev2022,Lukashchuk2024}; fiber optics and photonics \cite{Adany2009,Chong2006,Wise2008, Cao2021,Katz2011,Lihachev2022,Lukashchuk2024,Snigirev2023,Kumar2024}; frequency domain reflectometry \cite{Saperstein2007,Delfyett2012}; attosecond pulse control \cite{Hofmann2019,Yusoff2024,Vincenti2012}; qubit control \cite{Hawkins2012,Kuzmanovic2024,Osullivan2022}; \sout{digital computing \cite{Delfyett2012};} wireless communication \cite{Ouyang2016,Li2021} and biological/medical imaging \cite{Malinovskaya2007a,Malinovskaya2007b,Mamou2008,Niederriter2017,BoerOCT2017}. Chirped pulses are widely utilized for their ability to be compressed and expanded without altering their frequency bandwidth. This allows for ultrashort intense pulses to be safely amplified without changing their spectrum. This also allows for the creation of long pulses with large frequency bandwidths which then provide long range high resolution measurements that are robust against dispersion. These advantages are what have made chirped pulses so versatile. In all these applications, it is imperative to have precise control over the propagation of a chirped pulse through a given scattering environment, particularly its time delay.

In this paper we focus on understanding and controlling the time delay of linearly chirped pulses. In particular we are concerned with \textit{linear} pulse propagation through a non-Hermitian \rev{lossy} dispersive system where the incident waves can be related to the output waves via the transmission or reflection matrix \cite{Popoff2010,Chen2022}; defined as sub-matrices \sout{(or coefficients if $M\leq 2$)} of the \sout{$M\times M$} scattering matrix (S) \sout{where $M$ is the total number of scattering channels}. \rev{This work strives to create a connection between the behavior of chirped pulses in non-Hermitian scattering systems and a version of Wigner-Smith time delay \cite{Eisenbud1948,Wigner1955,Smith1960} adapted for non-Hermitian settings \cite{Chen2021,Chen2022}. In this context, we use the term non-Hermitian to indicate that we are working with open dissipative systems, as established in the photonics and wave-scattering literature \cite{Elganainy2018,Ozdemir2019,Ashida2020,Xue2026}.}

In this paper we use an extension of Wigner-Smith time delay \cite{Eisenbud1948,Wigner1955,Smith1960} to describe how long a wave lingers in a non-Hermitian scattering system. Wigner-Smith time delay was originally defined in the context of unitary nuclear scattering \sout{as $\tau_W(E) = -\frac{i}{M}\frac{d}{dE}\text{ln[det}S(E)]$ where $E$ is energy. This concept} \rev{and} was later generalized to include non-unitary dispersive scattering systems in the context of classical electromagnetic waves \cite{Chen2021,Chen2022}. We use this extension, referred to as complex time delay (CTD) \rev{or non-Hermitian Wigner Smith Time delay}, to predict the time shift of a chirped electromagnetic pulse after it travels through a resonant scattering system. CTD is expressed as,
%  \begin{align}
%     \tau_T = -i\frac{\partial}{\partial \omega}\text{ln[T}(\omega+i\alpha)] \qquad\qquad \tau_R = -i\frac{\partial}{\partial \omega}\text{ln[R}(\omega+i\alpha)]
%     \label{Eqn: CTDs}
% \end{align}
  \begin{align}
     \tau = \frac{-i}{2\pi}\frac{\partial}{\partial f}\text{ln[}\rho(f)]=\text{Re}[\tau]+i\text{Im}[\tau]
     \label{Eqn: CTDs}
 \end{align}
where $\tau$ is \sout{transmission/reflection } \rev{complex} time delay, $\rho$  \rev{represents either $det[S]/M$ (Wigner-Smith), or a} scattering matrix transmission (T) or reflection (R) coefficient of $S$, $f$ is frequency,  \rev{and $M$ is the number of scattering channels}. This generalization of time delay is complex due to the sub-unitary and dispersive nature of  \rev{$det[S]$ and} the matrix elements \cite{Asano2016,Chen2021,Hougne21,Chen2022,Huang22,Shaibe2025}. This can be seen by representing complex $\rho(f)$ in terms of a frequency dependent magnitude and phase. The real and imaginary parts of these delays can be positive, negative or zero. \rev{The use of CTD is general in the sense that it requires no symmetries of the Hamiltonian (e.g. Parity-Time \cite{Bender98}) or scattering matrix (e.g. reflection symmetry \cite{Muga21}).  We have established that CTD is equally valid in the context of both reciprocal and non-reciprocal scattering systems \cite{Shaibe2025,Shaibe2025PRR}.} \rev{Wigner-Smith time delay is one specific case of the Generalized Wigner-Smith Operator (GWSO) which has been developed as a model for scattering control with a wide range of applications directed towards non-Hermitian systems \cite{Ambichl17,Bliokh2025,Byrnes2025,Byrnes2025_2}.}

There has been considerable effort to understand the physical meaning behind negative real time delay, or fast light, measured with Gaussian pulses\cite{garrett1970,Crisp1971,chu1982,boyd2002, gehring2006,Bortolozzo2010}. This phenomenon occurs when the group velocity of light is either greater than the speed of light in vacuum, or negative. This happens, for example, in scattering systems near/at a resonance where there is a large amount of anomalous dispersion \cite{boyd2002}. These cases have been observed experimentally \cite{stenner2003,gehring2006} and are both interpreted as the result of extreme nonuniform attenuation of the Fourier components of the pulse as it travels through the anomalously dispersive medium. Note that the general pulse shape is not distorted as long as the frequency bandwidth of the pulse is less than the width of the resonance being excited in the system\cite{garrett1970,boyd2002}. The nonuniform change in the Fourier components leads to a shift in time of the pulse peak, creating the illusion that the pulse has left the system before it has entered \cite{Crisp1971,boyd2002}. This does not violate causality since there is a distinction between group velocity and information velocity as explained by Sommerfeld and Brillouin\cite{Sommerfeld1907,Brillouin1914,Brillouin1960} and experimentally demonstrated in Ref. \cite{stenner2003}. There has also been recent work on interpreting negative time delay in quantum mechanical systems, where in Refs.\cite{Thompson_2025,Angulo2026} they explore the time delay of a photon traveling through an atom cloud. %They found that the time a photon spends as an atomic excitation is directly related to the group delay and Wigner-Smith time delay, and interpreted negative time delay as the result of quantum interference that yields an anomalous weak value.  

Imaginary time delay has received considerably less attention than negative time delay/fast light. Imaginary transmission time delay was first properly predicted to correspond to a center frequency shift in a Gaussian pulse in Ref.\ \cite{Asano2016}, and experimentally confirmed in Ref.\ \cite{Giovannelli2025}. Imaginary time delay occurs for the same reason that negative time delay does: it is the result of nonuniform distortion of the Fourier components as the pulse travels through an anomalously dispersive linear medium. 

%It was shown in Refs.\, \cite{Asano2016, Giovannelli2025} that the imaginary part of complex time delay is directly proportional to a center frequency shift in a Gaussian pulse. In this paper we will show that not only is this true for chirped pulses but that the imaginary part also plays a significant role in the time shift of the peak of the pulse.

\rev{In previous work \cite{Asano2016,Giovannelli2025}, the relationship established between CTD derived from frequency-domain scattering parameters, and the time-domain pulse propagation properties, were only established for Gaussian wave packets.} We now wish to generalize \sout{the} \rev{these} complex time delay results \sout{presented in Refs.\ \cite{Asano2016, Giovannelli2025}} to chirped pulses, which are of great interest to many different fields and applications. Here we theoretically and experimentally demonstrate how the properties of a chirped pulse change as it travels through a linear \rev{non-Hermitian} dispersive scattering system. We also show how these changes in pulse propagation properties can be accurately predicted with only knowledge of the complex transmission (or reflection) coefficient as a function of frequency. \rev{Further} \sout{Specifically}, we analyze how the peak of the chirped pulse shifts in time as well as how its center frequency changes \rev{after it travels through a non-Hermitian scattering system}. \rev{We then demonstrate that these shifts are directly related to the corresponding CTD (reflection or transmission).} \rev{Specifically}, we confirm that the frequency shifts are directly proportional to imaginary time delay in a manner similar to that shown in Refs.\ \cite{Asano2016,Giovannelli2025}. We also show that the peak time shifts of a chirped pulse depend not only on the real part of complex time delay but also in a non-trivial way on the imaginary part.

%A chirped pulse is one in which the carrier frequency of the pulse is swept during the time the pulse has non-zero amplitude.  Frequency chirping is an alternative way to explore dispersion, and therefore will modify the CTD from their values in Eqn. 2-3 in \cite{Giovannelli2025}.  There is widespread use of chirped pulses in radar, lidar, etc.  We consider simple linear-in-time chirping, either positive or negative, in this work.  One of the main results is that these pulses have complex time delays that can be tailored beyond the values given in Ref.\,\cite{Asano2016,Giovannelli2025}. 

%The chirp rate is defined as $\Omega'$, and will be presented in units of GHz/ns in this paper.  

%%%%%%%%%%%%%%%%%%%%%%% Theory %%%%%%%%%%%%%%%%%%%%%%%%
\section{Chirped pulse properties and complex time delay}
The predicted results connecting complex time delay and chirped pulse properties can be derived following a process similar to that utilized in Refs.~\cite{Asano2016,Giovannelli2025, Cao2003}.  The details of this calculation can be found in the supplementary materials. We define a linearly chirped pulse to be the following:
% \begin{align}
%     E_i(\omega)\propto \text{exp}\left[\frac{-(\omega-\omega_o)^2}{2(i\Omega'+\delta_t^{-2})}\right]\
%     \label{Eqn: FD_InputSignal}
% \end{align}
\begin{align}
    E_i(t)\propto \text{exp}\left[-i2\pi f_0t-(i\Omega'+\delta_t^{-2})\frac{t^2}{2}\right]\
    \label{Eqn: TD_InputSignal}
\end{align}
where $f_0$ is the center frequency of the pulse, $\Omega^{\prime}$ is the linear chirp rate and $\delta_t=\frac{\delta_T}{2\sqrt{2\text{ln2}}}$ where $\delta_T$ is the full width at half maximum of the amplitude of the pulse in the time domain. The frequency bandwidth of the chirped pulse is subsequently defined as,
\begin{align}
    \delta_f=\frac{1}{\pi}\sqrt{2\text{ln}2(\Omega'^2\delta_t^2+\delta_t^{-2})}
    \label{Eqn: Bandwidth}
\end{align}
where $\delta_f$ is the FWHM of the chirped pulse amplitude in the frequency domain in Hz. Note that in this formulation of a chirped Gaussian pulse, the chirp rate is bounded by the chosen frequency bandwidth of the pulse,
\begin{align}
    \Omega'\leq\frac{(\pi\delta_f)^2}{4\text{ln}2}
\end{align}
The explanation for this bound is contained in section 2 of the supplementary materials.

We consider the case of a pulse propagating through a linear dispersive system for which the pulse frequency bandwidth (Eq.\,(\ref{Eqn: Bandwidth})) is much smaller than the 3 dB linewidth of the chosen resonance of the scattering system.  
%\textcolor{red}{We emphasize that our pulse only depends on time, so when we refer to the pulse ``propagating'' in this paper, we are referring to how the pulse is propagating in time-- not space.$\Rightarrow$} 
Note that we treat the scattering system in terms of its dispersive overall transmission/refection coefficient, rather than as a continuous extended medium in space. We define the shifts in frequency and time to be the shifts in frequency of the peak of the power spectral density, and shifts in time of the peak of the pulse power, respectively.  These are directly related to the CTD of the scattering system. The shift in center frequency is predicted to be,
\begin{align}
    D_f=-\tilde{\Delta}^2\text{Im}[\tau]
    \label{Eqn: Df}
\end{align}
where $D_f$ is the shift in center frequency in Hz and $\tilde{\Delta}=\frac{1}{2\pi}\sqrt{\Omega'^2\delta_t^2+\delta_t^{-2}} =\frac{\delta_f}{2\sqrt{2 \text{ln}2}}$ is proportional to the bandwidth of the chirped pulse. The time shift in the output pulse is found to be,
\begin{align}
    D_t=\text{Re}[\tau]-\Omega'\delta_t^2\text{Im}[\tau]
    \label{Eqn: Dt}
\end{align}
\noindent where $D_t$ is the predicted time shift of the chirped pulse which, surprisingly, depends on the imaginary part of time delay, in addition to the real part, weighted by a factor of the chirp time-bandwidth product $\Omega'\delta_t^2$. This allows for an extraordinary degree of control of $D_t$, as demonstrated in Fig.\,\ref{Transmission_Reflection: Dt0}, where we achieve the $D_t=0$ condition over a broad range of pulse center frequencies.

\begin{figure*}[ht!]
    \centering
    \includegraphics[width=\textwidth, trim=0cm 3.5cm 0cm 3cm, clip]{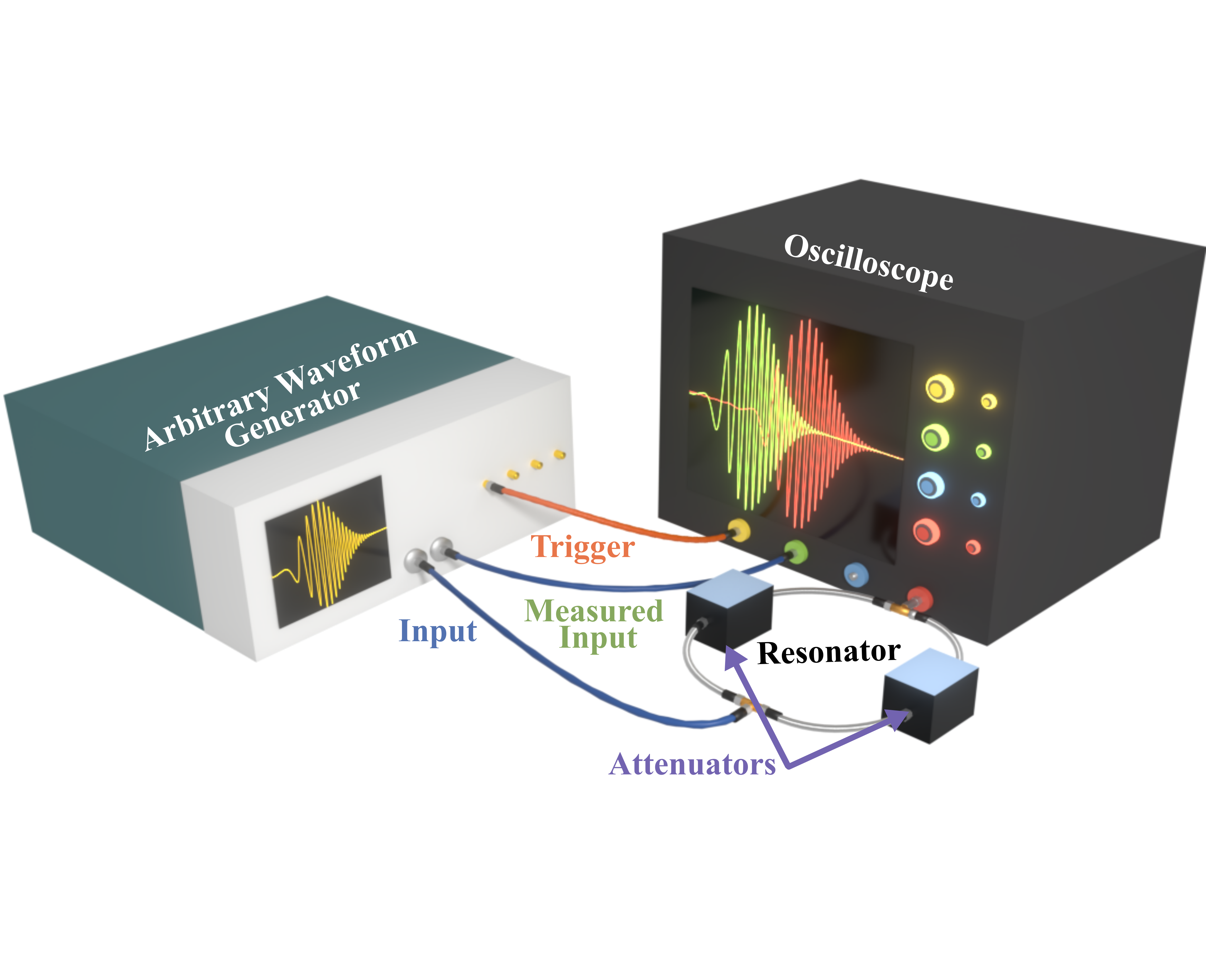}
    \caption{\rev{Schematic of time domain experiment setup.}\sout{(a) Schematic of time domain experiment setup.} The ring resonator depicted is the one that was used for the transmission case. \sout{(b) Complex time delays and transmission magnitude (in dashed orange) for the resonator depicted in (a). The red and dark purple solid lines respectively correspond to the real and imaginary parts of transmission time delay ($\tau_T$).}}
    \label{Experiment}
\end{figure*}

%The only restrictions on $D_t$ and $D_f$ are those imposed by the specific scattering system, equipment, and the requirement that the chirp is bounded $|\Omega'|\leq(\pi\delta_f)^2/4\text{ln}2$ for a given $\delta_f$, which is the result of the frequency bandwidth ($\delta_f)$ of the chirped pulse being defined as,
%\begin{align}
%    \delta_f=\frac{1}{\pi}\sqrt{2\text{ln}2(\Omega'^2\delta_t^2+\delta_t^{-2})}
%    \label{Eqn: Bandwidth}
%\end{align}
%where $\delta_f$ is the FWHM of the chirped pulse in the frequency domain in Hz. If one has an analytical expression for their transmission coefficient, they can derive concrete bounds for $D_t$ and $D_f$ using the weak-measurement formalism method laid out in Ref.\,\cite{Asano2016}.

% One can also deduce upper limits on the magnitude of time and frequency shifts of the chirped pulse...

%%%%%%%%%%%%%%%%%%%%%%% Experiment %%%%%%%%%%%%%%%%%%%%%%%%%%

\section{Experiment}
The time domain experiment setup is depicted in Fig.\,\ref{Experiment}, where we use a signal generator to create a chirped pulse that is sent through the resonant system and then measured on an oscilloscope. In more detail, a chirped pulse is produced using a 50 GSa/s Tektronix model AWG70001B arbitrary waveform generator (AWG). The time domain expression used to define the chirped pulse is given by the imaginary part of Eq.\,(\ref{Eqn: TD_InputSignal}). The two output ports on the AWG (1+, 1-) produce nominally identical pulses that are sent through twin 33.02 cm-long coaxial cables. Port 1- is connected directly to channel 2 of a Keysight/Infiniium model UXR0104A 25-GHz bandwidth real-time digital sampling oscilloscope (DSO), which captures the "measured input pulse".  Port 1+ of the AWG is connected to the resonator. The output pulse from the resonator is then measured on channel 4 of the oscilloscope. Lastly, the orange cable connects the AWG marker channel (M1+) to channel 1 of the oscilloscope and acts as the trigger signal. We note that Port 1- is the differential complement of Port 1+. The pulses coming out of these ports need to be identical in order to have a valid comparison between the pulse that travels through the "measured input cable" and the pulse that travels through the whole system. We meet this condition by simply multiplying the output signal from Port 1- by -1.

\begin{figure*}[ht!]
    \centering
    \includegraphics[width=\textwidth]{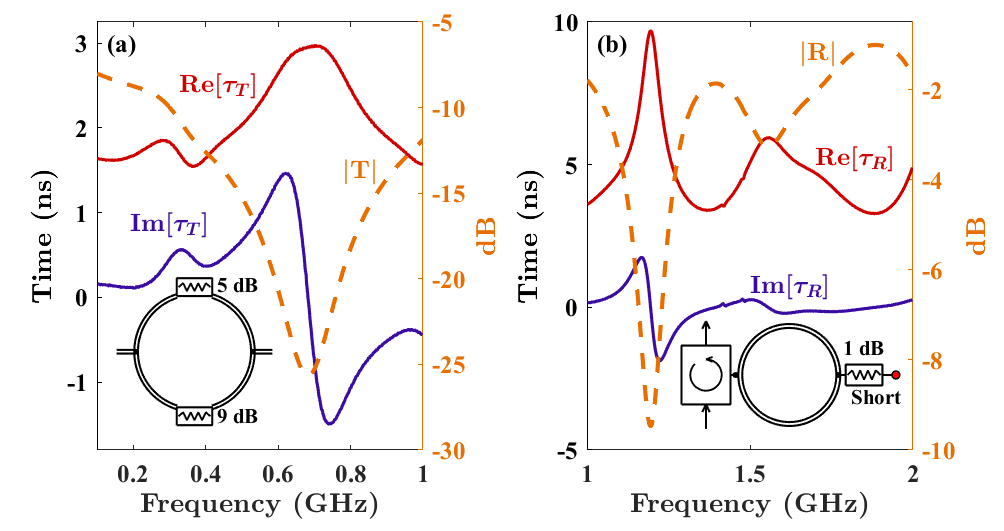}
    \caption{\rev{(a) The transmission scattering parameter (dashed orange, right axis), real (red) and imaginary (purple) parts of the transmission complex time delay (left axis). A schematic of the transmission resonator is depicted in the lower left. (b) The reflection scattering parameter (dashed orange) and real (red) and imaginary (purple) parts of reflection complex time delay.  A schematic of the resonator used for the reflection case is depicted in the lower right, where the block containing a circular arrow is a circulator.}}
    \label{Transmission SParam and Time Delays}
\end{figure*}
\begin{figure*}[ht!]
    \centering
    \includegraphics[width=\textwidth]{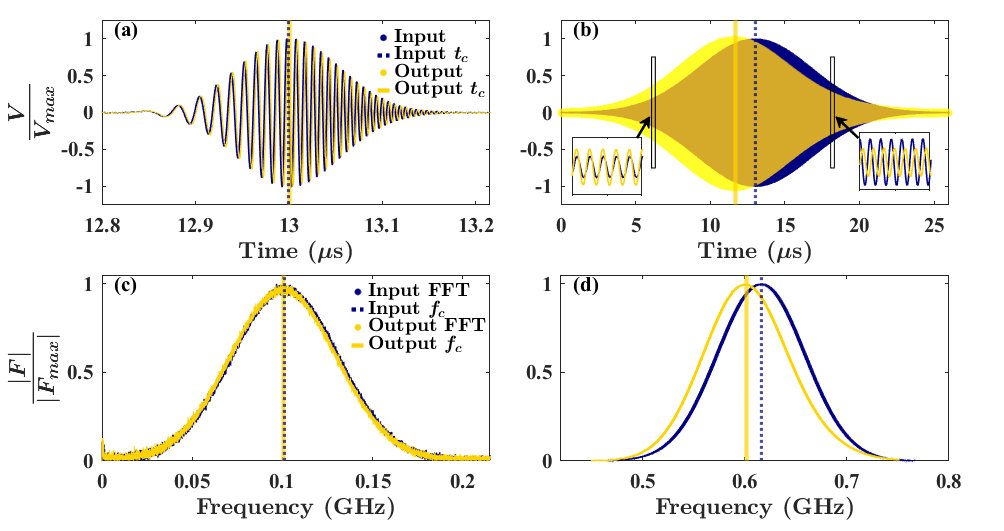}
    \caption{
     \rev{Panels (a) and (b) are examples of the time domain chirped pulses sent through the resonator depicted in Fig.\,\ref{Experiment}. The pulses sent into the resonator are plotted in blue and the output pulses are plotted in yellow. The transmission times are calculated as the mean of the envelope of the pulse $\left(t_c = \int |V(t)|^2t\:dt/\int |V(t)|^2\:dt\right)$ and are shown as corresponding vertical lines. The solid yellow vertical line is the transmission time of the output pulse and the dashed blue vertical line is the transmission time of the input pulse. The Fourier transforms (FFTs) of the pulses in (a) and (b) are depicted in panels (c) and (d) respectively. The calculated center frequencies $\left(f_c = \int |F(f)|^2f\:df/\int |F(f)|^2\:df\right)$ are marked on the plots as vertical lines: solid yellow for the output pulse center frequency and dashed blue for the input pulse center frequency.
     Panels (a) and (c) are an example of chirped pulse with an extremely small magnitude time and frequency shift (from data set depicted in Fig.\,\ref{Transmission_Reflection: Dt0}(a)). Panels (b) and (d) are an example of a pulse with a large time and center frequency shift from the dataset depicted in Fig.\,\ref{mainresults}(b). The insets in (b) show detailed oscillations of the pulses in equal-width (10 ns) time windows, illustrating the chirped nature of the pulse.}
    \sout{Examples of the normalized chirped pulses sent through the resonator depicted in Fig.\,\ref{Experiment}. The pulses sent into the resonator are plotted in blue and the output pulses are plotted in yellow. The measured center frequencies ($f_c)$ and transmission times ($t_c$) are shown as corresponding vertical lines. The insets in (b) show detailed oscillations of the pulses in equal-width (10 ns) time windows, illustrating the chirped nature of the pulse.}}
    \label{TDFD_Example}
\end{figure*}

In a separate experiment, the complex transmission $T(f)$ and reflection $R(f)$ coefficients of the ring resonator were measured using a Keysight N5242A network analyzer (PNA-X) that is calibrated using a Keysight N4691-60001 Electronic Calibration kit at the resonator's connection points. The measured transmission coefficient magnitude shown in Fig.\,\ref{Transmission SParam and Time Delays}(a) is taken over the frequency range 0.1 GHz to 1 GHz with a frequency step size of 1.331 MHz. The reflection coefficient magnitude shown in Fig.\,\ref{Transmission SParam and Time Delays}(b) was measured over the frequency range of 1 GHz to 2 GHz with a frequency step size of 1.66 MHz.  The corresponding real and imaginary parts of complex transmission and reflection time delay calculated using Eq. (\ref{Eqn: CTDs}), are shown as solid lines in Fig.\ \ref{Transmission SParam and Time Delays}.

\begin{figure*}[ht!]
    \centering
    \includegraphics[width=\textwidth]{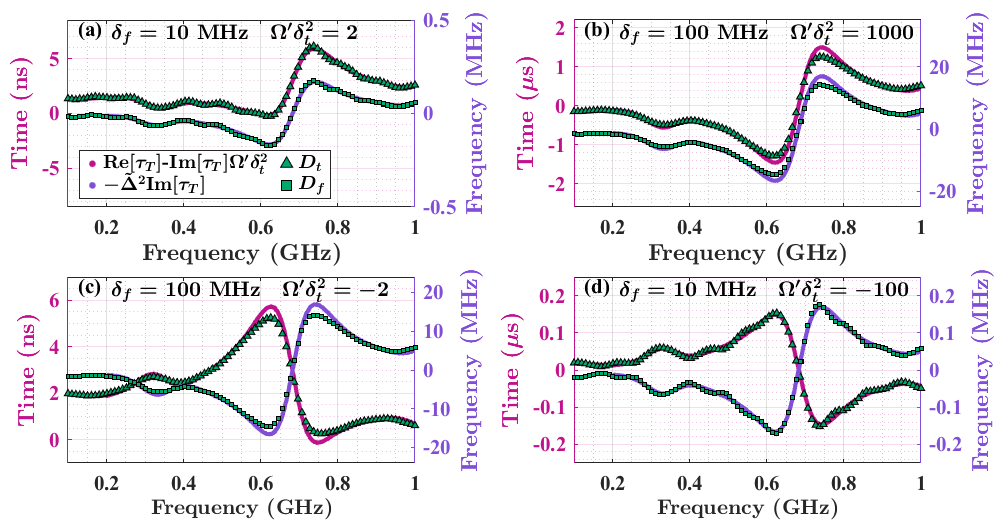}
    \caption{Summary of measured and expected transmission time delays and frequency shifts for chirped pulses having various $\Omega'\delta_t^2$ and $\delta_f$ values, as a function of center frequency. The predicted time shift is plotted in magenta and the measured pulse time shift are the green triangles. The predicted frequency shift is in light purple and the measured pulse frequency shift is plotted as green squares. The chirp rates and pulse widths used are as follows: (a) $\Omega'=0.2848$\,$\frac{\text{MHz}\,\text{Rad}}{\text{ns}}$ and $\delta_t=83.8$\,ns, (b) $\Omega'=0.0712$\,$\frac{\text{MHz}\,\text{Rad}}{\text{ns}}$ and $\delta_t=3747.82$\,ns, (c) $\Omega'=-28.48$\,$\frac{\text{MHz}\,\text{Rad}}{\text{ns}}$ and $\delta_t=8.38$\,ns, (d) $\Omega'=-0.0071$\,$\frac{\text{MHz}\,\text{Rad}}{\text{ns}}$ and $\delta_t=3748$\,ns}.
    \label{mainresults}
\end{figure*}

The transmission and reflection measurements were performed using the same experimental setup, but with two different resonators. Different resonators were used in the two cases to establish wide resonances of comparable size, which then provided similarly sized large $D_t$ and $D_f$ values for transmission and reflection, which are shown in Figs.\,\ref{mainresults} and \ref{SParam_ReflecTime}, respectively. The transmission resonance is situated at 0.684 GHz and has a 3-dB frequency bandwidth of about 117 MHz. The reflection resonance is centered at 1.19 GHz with a 3-dB bandwidth of 82 MHz.

The resonator used in the transmission case was a \rev{dispersive} microwave ring resonator \cite{Walt2013,Chen2022} consisting of 4 coaxial cables and 2 identical Narda (model 4745-69) step attenuators, as shown in the lower left of Fig.\,\ref{Transmission SParam and Time Delays}(a). \rev{These attenuators create additional loss and dissipation in order to widen the resonant modes of the system}. Two of the cables are each 15.24 cm-long and are attached to either side of a step attenuator set at 5 dB. The other two cables are each 7.62 cm-long and are attached to either side of the other step attenuator set at 9 dB. The ring is completed with two `tee` junctions, creating a 2-port network. \rev{This resonator is reciprocal, allowing us to say $\tau_{12}=\tau_{21}$, hence we only refer to one transmission coefficient (T) when describing the system. The symmetry in this system does not play a pivotal role in the theory or results we present. The simple experimental design was chosen to allow for clear interpretation of the results. See Refs.\,\cite{Muga21,Bender98} for how symmetry can play a role in non-Hermitian scattering systems}

The ring resonator used for the reflection case is depicted on the lower right of Fig.\,\ref{Transmission SParam and Time Delays}(b). It is composed of a 1-2 GHz Pasternack model PE83CR003 microwave circulator that is connected through a `tee` junction to two identical 12.7 cm-long coaxial cables forming a ring with a Narda (model 4745-69) attenuator set to 1 dB connected with a `tee` junction on the other side. The opposing port on the attenuator was connected to a short circuit to encourage the wave to make a round trip through the attenuator before returning to the ring. One port of the circulator is the input, while the output was connected to a male-to-male adapter that was directly connected to the oscilloscope. \rev{All of the elements in this resonator are lossy and add dissipation to the system.}

%%%%%%%%%%%%%%%%%%%%%%% Discussion %%%%%%%%%%%%%%%%%%%%%%%%%%%

\section{Data and Discussion}

\begin{figure*}[ht!]
    \centering
    \includegraphics[width=\textwidth]{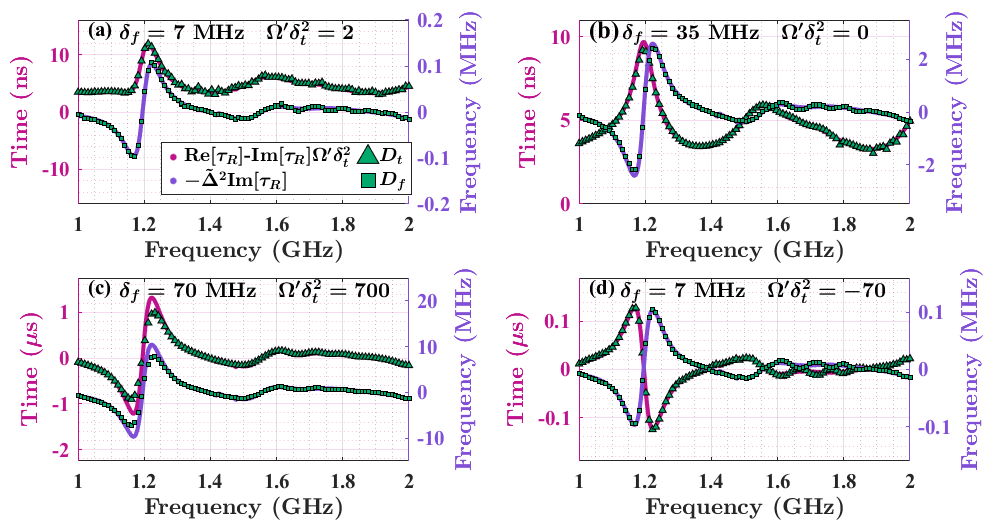}
    
    \caption{\sout{(a) Schematic of reflection resonator. (b) The red and dark purple solid curves are the real and imaginary parts of reflection time delay respectively. The orange dashed curve corresponds to the magnitude of the reflection coefficient for the resonator in (a).} Summary of measured and expected time shifts and frequency shifts for chirped pulses after reflection from a ring resonator.  The predicted center frequency shifts are shown in in light purple, and predicted time shifts in magenta with the measured frequency shifts represented as green rectangles and time shifts represented as green triangles. This is done for various chirped pulse frequency bandwidths ($\delta_f$) and $\Omega'\delta_t^2$ values.  The chirp rates and pulse widths used are as follows: (a) $\Omega'=0.1395$\,$\frac{\text{MHz}\,\text{Rad}}{\text{ns}}$ and $\delta_t=120$\,ns, (b) $\Omega'=0$\,$\frac{\text{MHz}\,\text{Rad}}{\text{ns}}$ and $\delta_t=10.71$\,ns, (c) $\Omega'=0.0498$\,$\frac{\text{MHz}\,\text{Rad}}{\text{ns}}$ and $\delta_t=3747.80$\,ns, (d) $\Omega'=-0.005$\,$\frac{\text{MHz}\,\text{Rad}}{\text{ns}}$ and $\delta_t=3747.87$\,ns.}
    \label{SParam_ReflecTime}
\end{figure*}

Figure \ref{TDFD_Example} shows normalized raw transmitted pulse data of a few representative measured input and output pulses, their corresponding Fourier transforms, and the time/frequency shifts extracted from the data. The input pulses are in dark blue and the corresponding outputs are in yellow. Panels (a) and (c) of Fig. \ref{TDFD_Example} correspond to a pulse with center frequency $f_c = 0.1013$ GHz and a chirp rate of $\Omega' = 3.39$\,$\frac{\text{MHz}\,\text{Rad}}{\text{ns}}$. This pulse has a time width ($\delta_t$) of 54.89 ns, a frequency bandwidth ($\delta_f$) of 70 MHz, and a chirp factor ($\Omega'\delta_t^2$) of 10.20. This serves as an example of a pulse with both a minimum transmission time shift ($D_t=0.10\pm 0.03$ ns) and a minimum center frequency shift ($D_f=-0.8479\pm 0.0140$ MHz). (The uncertainty estimates are detailed in the Supp. Mat.)  Panels (b) and (c) of Fig. \ref{TDFD_Example} correspond to a pulse from the data shown in Fig.\ref{mainresults}\,(b) with a center frequency of $0.6157$ GHz, which situates it right near peak resonance. The insets on either side of the pulse are zoomed-in regions of detailed oscillations to illustrate that the pulse is not just a Gaussian pulse but a chirped Gaussian pulse. The pulse has a chirp rate of $0.071$\,$\frac{\text{MHz}\,\text{Rad}}{\text{ns}}$, and a pulse length ($\delta_t$) of $3.75\,\mu$s. This is an example of a pulse that undergoes both a large negative time shift ($D_t=-1.2904\pm 0.0001\,\mu$s) and a large negative frequency shift ($D_f=-14.6576\pm0.0006$ MHz). The transmission times ($t_c$) and center frequencies ($f_c$) are depicted as vertical lines in Fig.\,\ref{TDFD_Example}, and are calculated using the first temporal moment of the pulse, as done in Ref.\cite{Giovannelli2025}. This is valid when working in the small bandwidth limit where the frequency bandwidth of the pulse is less than the 3-dB bandwidth of the excited resonance, hence higher-order terms in the phase and attenuation vs. frequency expressions can be neglected (see Supp. Mat. Section 1). The pulse in panels (a) and (c) has a frequency bandwidth of 70 MHz, and the pulse in panels (b) and (d) has a bandwidth of 100 MHz which are both smaller than the width of the transmission resonance (117 MHz).

\begin{figure*}[ht!]
    \centering
    \includegraphics[width=\textwidth]{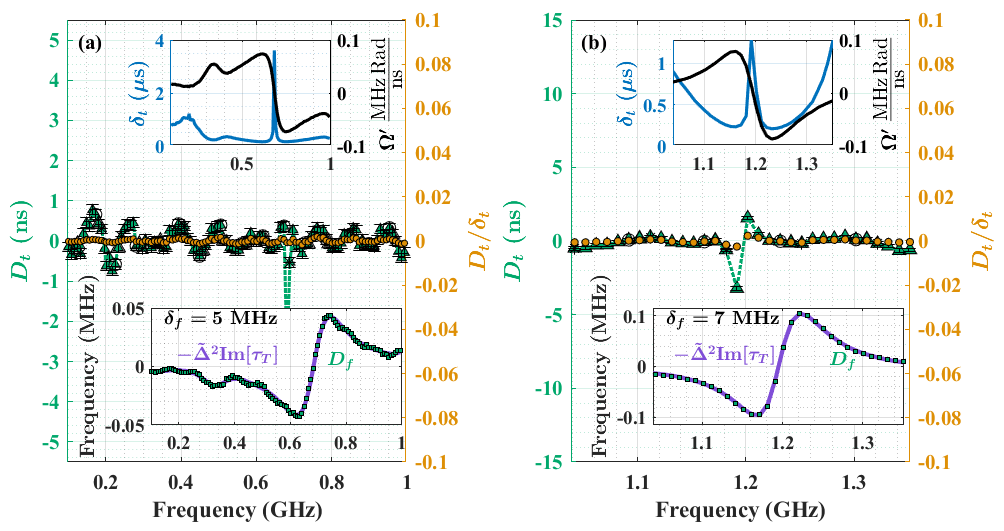}
    
    \caption{Results for the case where the chirped-pulse temporal shift $D_t$ is zeroed out by setting $\text{Re}[\tau]/\text{Im}[\tau]=\Omega'\delta_t^2$ at each choice of center frequency. Panels (a) and (b) are the results for transmission and reflection, respectively. The green triangles correspond to the measured chirped pulse transmission (a) and reflection (b) time shift (left axis). The light orange circles correspond to time-shift normalized to the width of the pulse (right axis). \sout{The real and imaginary parts of transmission (a) and reflection (b) time delay are plotted in red and dark purple respectively.} The lower insets in (a) and (b) show the predicted and measured center frequency shifts plotted in light purple and as green squares, respectively, along with the value of the common pulse bandwidth $\delta_f$.  The upper insets show the chosen values of $\delta_t$ and $\Omega'$ versus center frequency utilized to achieve $D_t=0$ for the chirped pulse.} 
    \label{Transmission_Reflection: Dt0}
\end{figure*}

The main results comparing the measured time and frequency shifts with the predictions made in Eqs.\,\ref{Eqn: Df}-\ref{Eqn: Dt} are shown in Figs.\,\ref{mainresults} and \ref{SParam_ReflecTime}, for transmission and reflection, respectively. The time-domain transmission data consists of 93 data points taken over the pulse center frequency range 0.1-1 GHz and the reflection time-domain data consists of 32 data points taken over 1-2 GHz. The symbols show the time-domain $D_t$ and $D_f$ measurements, while the solid lines show quantities constructed from frequency-domain data to test Eqs. (\ref{Eqn: Df}) and (\ref{Eqn: Dt}).  Shown are data from a variety of pulses with different bandwidths $\delta_f$ and chirp rates $\Omega'\delta_t^2$ (positive and negative).  From this we see that the predictions are consistent with the experimental results for a variety of pulse frequency bandwidths and  chirp parameter values. We see that the frequency shifts ($D_f$) scale with the pulse frequency bandwidth squared, as expected from Eq.\,\ref{Eqn: Df}. We also see that there is less agreement for large frequency bandwidths as they approach the 3-dB-linewidth of the resonance being excited. This is expected as this is where the theory starts to break down (see Supp. Mat. Section 1). This was the main motivation for designing resonators to have wide modes so that we could achieve such large frequency shifts. 
%\textcolor{blue}{I suggesting removing text from here ...}\rev{When the frequency bandwidth of the pulse is too large the output becomes incredibly distorted. Additionally, a larger frequency bandwidth pulse means its Fourier components have a larger range. When there are too many Fourier components that are outside the influence of the absorptive resonance, they outweigh the components distorted by the resonance, essentially nullifying the clean shifting behavior the resonance enacts on the pulse. What is left is a visually distorted pulse that is effectively missing a chunk of the original pulse's Fourier components. \textcolor{blue}{...to here} 
\rev{As the pulse frequency bandwidth approaches the width of the resonance, the measured time and frequency shifts decrease, eventually approaching zero for an isolated mode. This case is demonstrated in Fig. 3(b), (d) of the supplementary materials of Ref.\,\cite{Giovannelli2025}. This behavior is also partially demonstrated in Fig.\,\ref{mainresults}(c) and Fig.\,\ref{SParam_ReflecTime}(c) where we see that at larger bandwidths the measured pulse shifts decrease in amplitude in comparison to the shifts predicted by frequency domain CTD.} 

Note that the case of an un-chirped pulse is included for reflected pulses in Fig.\,\ref{SParam_ReflecTime}(b).  \rev{This case was not reported in Ref.\,\cite{Giovannelli2025}, and is demonstrated for the first time here.}  \sout{; the un-chirped case for transmission was already experimentally demonstrated in Ref.\,\cite{Giovannelli2025} so it is omitted from this paper.} Additional data for the pulse reflection case is included in Supp. Mat. Section 3.

It is clear from the data that the time shift of the pulse is heavily dependent on the magnitude of the dimensionless chirp parameter $\Omega'\delta_t^2$. We see that in both transmission and reflection, the time shift jumps from being on the scale of nanoseconds to microseconds, scaling with $\Omega'\delta_t^2$. We note that in the limit of large $\Omega'\delta_t^2$, and/or large $\text{Im}[\tau]$, $D_t$ converges to a frequency dependence that is identical to that of $D_f$ (modulo a sign). This makes sense when looking at Eq.\,\ref{Eqn: Dt} where the imaginary part of time delay is expected to dominate the time shift in the chirped pulse when $\Omega'\delta_t^2$ is large. This is also true for the negative chirp case, except for a minus sign, following the sign of the chirp rate (as demonstrated in Figs.\,\ref{mainresults}(c-d) and \,\ref{SParam_ReflecTime}(d)).

In Fig.\,\ref{Transmission_Reflection: Dt0} we demonstrate how with knowledge of complex time delay one can manipulate the chirped pulse properties such that the time shift of the pulse is zero over a broad range of center frequencies. Combining Eqs.\,\ref{Eqn: Dt} and \ref{Eqn: Bandwidth} allows one to choose unique values for $\delta_t$ and $\Omega'$, for a given pulse bandwidth $\delta_f$, such that $\text{Re}[\tau]/\text{Im}[\tau]=\Omega'\delta_t^2$, resulting in a zero time shift $D_t=0$ of the chirped pulse. \rev{(Note that one must choose new values of $\Omega'\delta_t^2$ at each pulse center frequency because $\text{Re}[\tau]$ and $\text{Im}[\tau]$ are frequency dependent.)} We tested this experimentally for both transmission and reflection, and the results are shown in Fig.\,\ref{Transmission_Reflection: Dt0}.  Shown there are \sout{$\text{Re}[\tau]$ and $\text{Im}[\tau]$ (red and blue solid lines) vs. frequency, as well as the resulting} \rev{the} time shifts $D_t$ (green symbols) over a broad range of pulse center frequency, for both transmission and reflection.   The average $D_t$ value over the bandwidth found for the transmission case in Fig.\,\ref{Transmission_Reflection: Dt0}(a) is -0.03$\pm 0.12$\,ns, which is a shift in time that is 0.0026\% of the pulse width. The average $D_t$ value found for the reflection case in Fig.\,\ref{Transmission_Reflection: Dt0}(b) is -0.14$\pm0.12$\,ns or a 0.0091\% temporal shift. We see that it is indeed possible to achieve $D_t=0$ for chirped pulses over a broad range of center frequencies. We found for pulses with a frequency bandwidth around 5 MHz, that the uncertainty in the time shift is around $\pm0.12$\,ns and the uncertainty in the calculated frequency shift is approximately  $\pm0.6$\,kHz. The error bars on the shift in time is shown in Fig.\,\ref{Transmission_Reflection: Dt0}, whereas we do not place error bars on the frequency shift measurements because the uncertainty is smaller than the size of the data symbol.

Achieving $D_t \approx$ 0 was significantly more difficult for the reflection case, where we were only able to achieve well behaved $D_t\approx0$ behavior near resonance. When looking off resonance at the imaginary part of reflection time delay (shown in Fig.\,\ref{Transmission_Reflection: Dt0}(b)) we see that there are multiple zero crossings and that in general the imaginary part tends to hover around zero. Since we are imposing $\Omega'\delta_t^2 = \text{Re}[\tau]/\text{Im}[\tau]$ we expect there to be divergences in chirp parameter when $\text{Im}[\tau]=0$. The full data set is located in Fig.\,4 of the Supp Mat, where we see sharp increases in $D_t$ where $\text{Im}[\tau]$=0. 
%Since there are so many zero crossings in the imaginary part of our reflection time delay, these divergences start to take over leading multiple divergences. 
Additionally, see Fig.\,3 in the Supp Mat where we perform this same study with pulses having considerably wider (50-70 MHz) frequency bandwidth. 

%The figure shows that it is indeed possible to achieve $D_t=0$ for chirped pulses over a broad range of center frequencies. In more detail, we see that the raw $D_t$ values in green are smoother and closer to zero for the cases where the bandwidth is large. However when looking at the normalized $D_t/\delta_t$ we see that it is significantly smoother and closer to zero for the smaller bandwidth case. This is the result of the $\delta_t$ values for the smaller bandwidth case being significantly larger (on average) than the larger bandwidth case but experiencing $D_t$ shifts on similar scales. The chosen $\delta_t$ and $\Omega'$ values are shown in the upper left of Fig.\,\ref{Transmission_Reflection: Dt0}.

%%%%%%%%%%%%%%%%%%%%%%% Conclusion %%%%%%%%%%%%%%%%%%%%%%%%%%%
\section{Conclusion}
%In this paper we expand on the work presented in Ref.\,\cite{Giovannelli2025,Asano2016} to chirped pulses. We demonstrate how complex time delay can be used to predict and control the properties of chirped pulses as they travel through a linear dispersive scattering system. This allows one to choose a chirp rate $\Omega'$, pulse duration $\delta_t$, and pulse bandwidth $\delta_f$ to maximize or minimize the pulse's shift in transmission (or reflection) time and center frequency.

%In terms of future work it would be interesting to see how these relationships change in more complex systems with many overlapping modes. It would also be interesting to extend these predictions to systems with more than 2 ports.

In this paper we have established a comprehensive theoretical and experimental framework connecting complex time delay to the properties of linearly chirped pulses traveling through \rev{linear and dispersive non-Hermitian} resonant scattering systems. Building upon the work in Refs.\,\cite{Giovannelli2025,Asano2016}, we have demonstrated that both the real and imaginary components of complex time delay play essential roles in determining chirped pulse behavior, with the imaginary part contributing not only to the observed frequency shifts but also to the temporal shifts weighted by the chirp parameter $\Omega'\delta_t^2$. \rev{As a result, this work broadens not only how we physically interpret complex time delay \cite{Chen2021,Chen2022} but also adds further motivation for why it is useful to generalize real time delay to non-Hermitian systems in the first place.}

The experimental validation for both cases of transmission and reflection, spanning a wide range of chirp rates (both positive and negative) and pulse frequency bandwidths, confirms the robustness of our theoretical predictions and their practical utility. This allows one to construct chirped pulses with predetermined temporal and spectral characteristics using only knowledge of the system's transmission or reflection coefficient as a function of frequency. This was demonstrated in Fig.\,\ref{Transmission_Reflection: Dt0} where we used the resonator's transmission/reflection coefficient to appropriately choose the chirp rate and pulse duration such that the time shift $D_t$ of the pulse is nullified over a wide frequency range. The ability to achieve near-zero broadband time shifts opens new possibilities for precision control in various applications ranging from pulse compression systems to quantum computing. This result also represents a new way to take advantage of the additional degrees of freedom present in chirped pulses, going beyond what is achievable with unchirped pulses alone.

Looking ahead, there are several interesting avenues for future work. Extension to multi-mode systems with overlapping resonances would test how our theory holds in more realistic environments where mode coupling and interference effects become important. Similarly, it would be interesting to see how our theory can be expanded to systems with more than two ports or for nonlinearly chirped pulses. Additionally, since in this paper we only used simple microwave ring resonators, it would be valuable to examine how a resonator's geometry or characteristics (i.e. coupling strength, loss, Q-factor, etc.) impact the achievable range for $D_t$ and $D_f$. Another open question would be how these results translate to the optical domain using systems such as an optical ring resonator or a whispering gallery mode resonator.

\begin{flushleft}
\textbf{Funding.} This work was partially supported by NSF/RINGS under grant No. ECCS-2148318, ONR under grant N000142312507, and DARPA WARDEN under grant HR00112120021. 
\end{flushleft}

\begin{flushleft}
\textbf{Disclosures.} The authors declare no conflicts of interest.
\end{flushleft}

\bibliography{References}
\newpage

\appendix
\title{Chirped Pulse Analysis and Control in Non-Hermitian Scattering Systems using Complex Time Delay Supplementary Materials}

\author{Isabella L. Giovannelli,\authormark{1,*} Steven M. Anlage,\authormark{1} and Thomas M. Antonsen\authormark{2}}

\address{\authormark{1}Quantum Materials Center, Department of Physics University of Maryland, College Park, Maryland 20742, USA\\
\authormark{2}Department of Electrical and Computer Engineering University of Maryland, College Park, Maryland, 20742, USA
}

\email{\authormark{*}igiovann@umd.edu}

%%%%%%%%%%%%%%%%%%%%%%%%%%  body  %%%%%%%%%%%%%%%%%%%%%%%%%%
 \section{Derivation of predictions}
 A chirped pulse is defined by the following equation in the time domain.
\begin{align}
    E_i(t)\propto \text{exp}\left[-i\omega_0t-(i\Omega'+\delta_t^{-2})\frac{t^2}{2}\right]
    \label{TDInput}
\end{align}

First we Fourier transform this equation into the frequency domain,
\begin{align}
   E_i(\omega)&\propto\frac{1}{\sqrt{2\pi}}\int^{\infty}_{-\infty}E_i(t)\, e^{i\omega t}dt \\
    &\propto \frac{1}{\sqrt{2\pi}}\int^{\infty}_{-\infty}\text{exp}\left[-\left(\frac{i\Omega'}{2}+\frac{1}{2\delta_t^2}\right)t^2+\left(-i\omega_0+i\omega\right)t\right]dt 
\end{align}
To evaluate the integral we simply complete the square, leaving us with the following result,
\begin{align}
    E_i(\omega)\propto\text{exp}\left[\frac{-(\omega-\omega_o)^2}{2(i\Omega'+\delta_t^{-2})}\right]
    \label{FDInput}
\end{align}
To generate the form for the output pulse, we follow the same steps that we did in the Supp. Mat. of Ref. \cite{Giovannelli2025}. The highlights will be provided here for convenience. The output pulse will take the form
\begin{align}
    E_o(\omega)&\propto \text{exp}[-\kappa+i\phi]E_i(\omega)
\end{align}
where $\rho(\omega)=\text{exp}[-\kappa+i\phi]$ is either the transmission or reflection coefficient of a dispersive scattering system, and $\kappa(\omega)$ is the attenuation and $\phi(\omega)$ is the phase shift. Taylor expanding $\kappa$ and $\phi$ around center frequency $\omega_0$ we get,
\begin{align}
    \kappa = \kappa(\omega_0) + \left.\frac{d\kappa(\omega)}
    {d\omega}\right|_{\omega_0}(\omega-\omega_0)+\cdots  \label{eq: kappa}\\
    \phi = \phi(\omega_c) + \left.\frac{d\phi(\omega)}{d\omega}\right|_{\omega_0}(\omega-\omega_0)+\cdots \label{TaylorExpansions}
\end{align}
Assuming a pulse frequency bandwidth that is much smaller than the 3dB linewidth of the chosen resonance, we will only use up to the first derivative with respect to frequency, ignoring higher order terms. Using the definition of time delay $\tau$ defined in the main text, one can easily find that $\partial\phi/\partial\omega=\text{Re}[\tau]$ and $\partial\kappa/\partial\omega=\text{Im}[\tau]$. Thus,
\begin{align}
    E_o(\omega)&\propto E_i(\omega)\text{exp}[{(-\text{Im}[\tau]+i\text{Re}[\tau])(\omega-\omega_0)]}
\end{align}

Separating real and imaginary parts in the exponents,
\begin{align}
    E_o(\omega)&\propto \text{exp}\left[{\frac{-(\omega-\omega_o)^2}{2(i\Omega'+\delta_t^{-2})}}\right]\text{exp}[{(-\text{Im}[\tau]+i\text{Re}[\tau])(\omega-\omega_0)]} \\
    &\propto\text{exp}\left[\frac{-(\omega-\omega_0)^2(-i\Omega'+\delta_t^{-2})}{2(\Omega'^2+\delta_t^{-4})}\right]\cdot\text{exp}\left[-\text{Im}[\tau](\omega-\omega_0)\right]\cdot\text{exp}\left[i\text{Re}[\tau](\omega-\omega_0)\right] \\
    &\propto\textcolor{red}{\text{exp}\left[\frac{-(\omega-\omega_0)^2}{2(\delta_t^2\Omega'^2+\delta_t^{-2})}-\text{Im}[\tau](\omega-\omega_0)\right]}\cdot\text{exp}\left[i\left(\frac{(\omega-\omega_0)^2\Omega'}{2(\Omega'^2+\delta_t^{-4})}+\text{Re}[\tau](\omega-\omega_0)\right)\right]
\end{align}
Now we complete the square in the argument of the red expression, this leaves us with the following,
\begin{align}
    E_o(\omega)&\propto\text{exp}\left[-\frac{[(\omega-\omega_0)-(\textcolor{blue}{-\tilde{\Delta}^2\text{Im}[\tau]})]^2}{2\tilde{\Delta}^2}\right]\text{exp}\left[i\left(\frac{(\omega-\omega_0)^2\Omega'}{2(\Omega'^2+\delta_t^{-4})}+\text{Re}[\tau](\omega-\omega_0)\right)\right] \label{eq: E_out(omega)}
\end{align}
where $\tilde{\Delta}=\sqrt{\Omega'^2\delta_t^2+\delta_t^{-2}}$. Note that $\tilde{\Delta}$ is written in terms of Hz in the main text, here we are using angular frequency units. The frequency shift in the magnitude is highlighted in blue.

To find the corresponding time shift, we take the inverse Fourier transform of Eqn. \ref{eq: E_out(omega)},
\begin{align}
    E_o(t)\propto\frac{1}{\sqrt{2\pi}}\int^\infty_{-\infty}E_o(\omega)e^{-i\omega t} d\omega
\end{align}
This results in the following expression,
\begin{align}
   |E_o(t)|&\propto \text{exp}\left[-\frac{\left((\text{Re}[\tau]-t\right)\tilde{\Delta}-\tilde{\Delta}^2\text{Im}[\tau]\delta_t\Omega')^2}{2\tilde{\Delta}^2\left(1+\delta_t^4\Omega'^2\right)}\right]\\
           &\propto \text{exp}\left[-\frac{\left(t-\left(\textcolor{blue}{\text{Re}[\tau]-\text{Im}[\tau]\delta_t^2\Omega'}\right)\right)^2}{2\delta_t^2}\right]
\end{align}
where we see that the time shift of the magnitude of the signal is highlighted in blue.

Note that these calculations were done entirely assuming frequency is written as angular frequency. The corresponding variables used in the main text are all written in linear frequency (Hz) to make interpreting the experiment results easier.

\newpage

\section{Derivation of bounds on $\Omega'$}
\begin{figure*}[ht!]
    \centering
    \includegraphics[width=\textwidth]{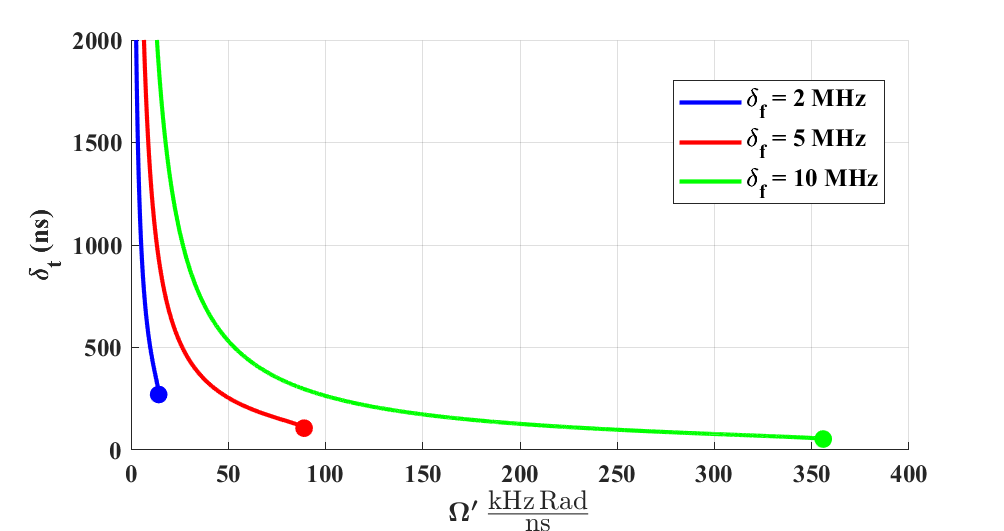}
    \caption{This is a plot of eqn.\,\ref{T} versus chirp rate ($\Omega'$) for several fixed chirped pulse frequency bandwidth values of 2 MHz (in blue), 5 MHz (in red), and 10 MHz (in green). The end points on each curve correspond to the maximum possible chirp value for a given frequency bandwidth.}
    \label{TvsC}
\end{figure*}
The frequency bandwidth of the chirped pulse (in Hz) is defined as,
\begin{align}
    \delta_f=\frac{1}{\pi}\sqrt{2\text{ln}2(\Omega'^2\delta_t^2+\delta_t^{-2})}
    \label{Eqn: Bandwidth}
\end{align}
Rearranging this equation we arrive at a quartic polynomial in terms of $\delta_t$,
\begin{align}
    \Omega'^2\delta_t^4-\frac{(\pi\delta_f)^2}{2\text{ln}2}\delta_t^2+1=0
    \label{Eqn: Polynomial}
\end{align}

There are 4 roots to this equation. Here we only consider the following root,
\begin{align}
    \delta_t=\sqrt{\frac{\frac{(\pi\delta_f)^2}{2\text{ln}2}+\sqrt{\left(\frac{(\pi\delta_f)^2}{2\text{ln}2}\right)^2-4\Omega'^2}}{2\Omega'^2}}
    \label{T}
\end{align}
Thus, we see that in order for $\delta_t\in\mathbb{R}$ we require,
\begin{align}
|\Omega'|\leq\frac{(\pi\delta_f)^2}{4\text{ln}2}
\label{ChirpLimit}
\end{align}
Eqn.\,\ref{T} is plotted versus chirp rate for various frequency bandwidth values in Fig.\,\ref{TvsC}. Here we see that the maximum chirp rate for a given frequency bandwidth increases with bandwidth as expected from eqn.\,\ref{ChirpLimit}. In general, we see that as $\Omega'$ increases, the pulse time width $\delta_t$ is bounded below as a result of the constraint on chirp rate (eqn.\,\ref{ChirpLimit}). When $\Omega'\rightarrow 0$ we see that $\delta_t\rightarrow\infty$, so in general the pulse time width can be made arbitrarily large for any finite pulse frequency bandwidth. It is also worth pointing out that there are an infinite number of combinations of $\delta_t$ and $\Omega'$ that give the same $\delta_f$ value. Thus, it is possible to have two chirped pulses with the same frequency bandwidth but with drastically different time shifts (see eqn.\,4 in the main text).

\newpage

\section{Additional Data}
\begin{figure*}[ht!]
    \centering
    \includegraphics[width=\textwidth]{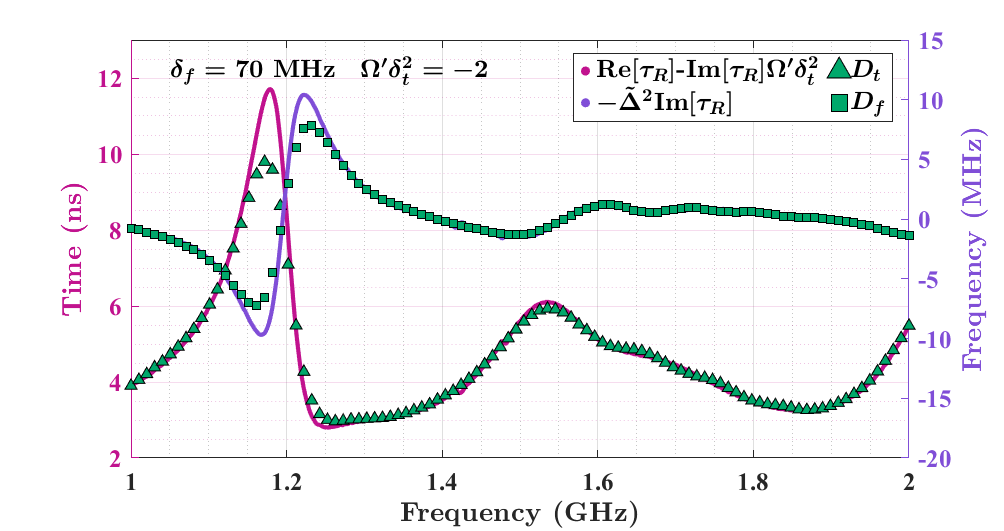}
    \caption{Additional chirped pulse reflection data comparing the predicted center frequency shifts in light purple, and predicted time shifts in magenta with the measured frequency shifts represented as green rectangles and time shifts represented as green triangles.  The chirp rate and temporal width of the pulse used is $\Omega'=0.1395$\,$\frac{\text{MHz}\,\text{Rad}}{\text{ns}}$ and $\delta_t=119.72$\,ns.}
    \label{AdditionalReflectionResults}
\end{figure*}

\begin{figure*}[ht!]
    \centering
    \includegraphics[width=\textwidth]{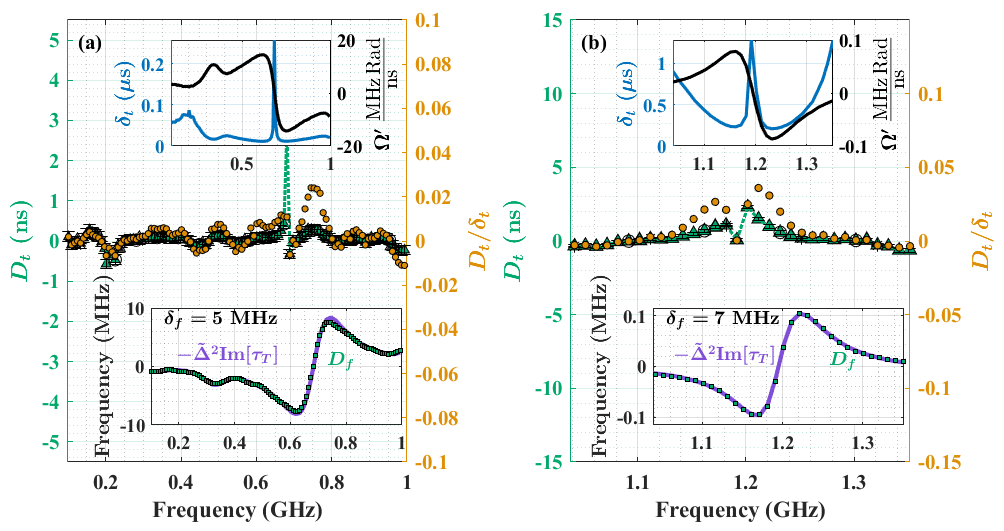}
    \caption{Additional $D_t=0$ data where larger frequency bandwidth pulses are used. Panels (a) and (b) are the results for transmission and reflection, respectively. The green triangles correspond to the measured pulse transmission time (left axis). The light orange circles correspond to $D_t/\delta_t$ (right axis). The real and imaginary parts of transmission (a) and reflection (b) time delay are plotted in red and dark purple, respectively. The lower insets in (a) and (b) show the predicted and measured pulse frequency shifts plotted in light purple and as green squares respectively.  The upper insets show the chosen values of pulse duration $\delta_t$ and chirp rate $\Omega'$ versus frequency utilized to achieve $D_t=0$ for the chirped pulse.} 
    \label{Transmission_Reflection_SuppMat: Dt0}
\end{figure*}

\begin{figure*}[ht!]
    \centering
    \includegraphics[width=\textwidth]{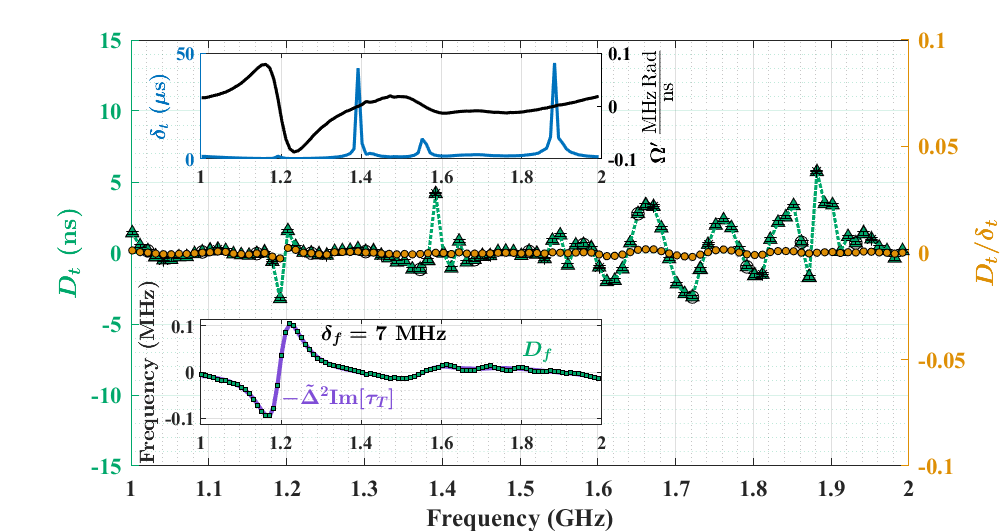}
    \caption{Additional chirped pulse reflection data showing the full data set of Fig. 6(b) in the main text. The green triangles correspond to the measured pulse transmission time (left axis). The orange circles correspond to $D_t/\delta_t$ (right axis). The real and imaginary parts of reflection time delay are plotted in red and dark purple respectively. The lower inset shows the predicted and measured pulse frequency shifts plotted in light purple and as green squares respectively.  The upper inset shows the chosen values of $\delta_t$ and $\Omega'$ versus frequency required to achieve $D_t=0$ for the chirped pulse.} 
    \label{Reflection_SuppMat_FullData: Dt0}
\end{figure*}

Figure\,\ref{AdditionalReflectionResults} contains additional chirped pulse reflection data that compliments the results shown in Fig.\,4 of the main text. Figure\,\ref{Transmission_Reflection_SuppMat: Dt0} contains data that is analogous to Fig.\,6 in the main text except for the case where the pulse has a frequency bandwidth of 70 MHz (transmission) and 50 MHz (reflection).
Figure\,\ref{Reflection_SuppMat_FullData: Dt0} is the full data set of Fig.\,6(b) in the main text. Here we show the multiple divergences for the expected chirp parameter $\Omega'\delta_t^2$ as the imaginary part of reflection time delay crosses through zero.

\clearpage

\section{Error Analysis}
\begin{figure*}[ht!]
    \centering
    \includegraphics[width=\textwidth]{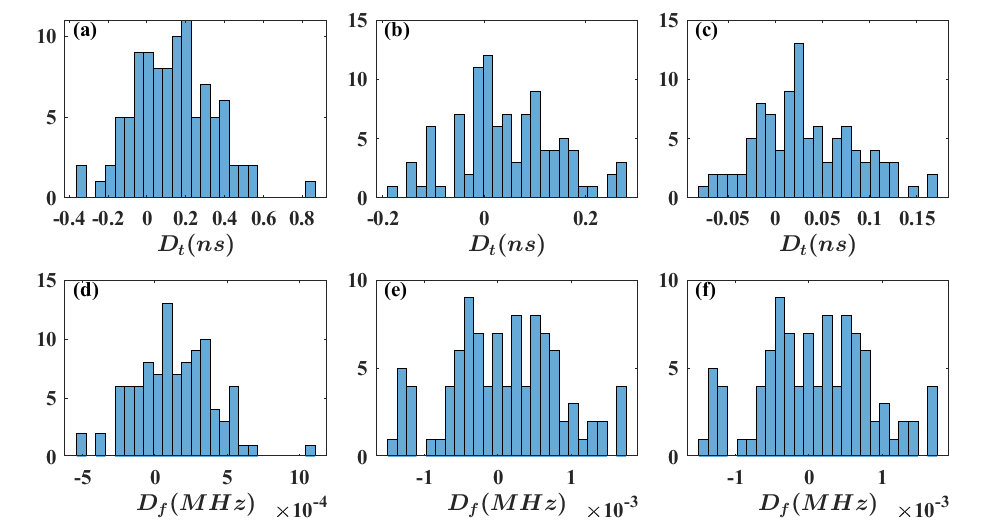}
    \caption{Time and frequency error analysis for the case where the input pulse has a frequency bandwidth of 5 MHz. Each histogram has 25 bins and 100 data points. The parameters for the plots are as follows: (a),(d) The pulses used have a chirp rate of $\Omega'$= 0.008 $\frac{\text{MHz}\,\text{Rad}}{\text{ns}}$ and a time width of $\delta_t$ = 1666\,ns. The mean and standard deviation for (a) are 0.13903\,ns and 0.2078\,ns respectively. For (d) the mean is 0.1384\,kHz and the standard deviation is 0.2732 kHz. (b),(e) The pulses have a chirp rate of 0.04 $\frac{\text{MHz}\,\text{Rad}}{\text{ns}}$ and a time width of 324\,ns. The histogram in (b) has a mean of 0.0442\,ns and a standard deviation of 0.1001\,ns. The analogous histogram (e) has a mean of 92.69\,Hz and standard deviation of 0.7642\,kHz. Lastly, for (c) and (f) the pulses used have a chirp rate of 0.08\,$\frac{\text{MHz}\,\text{Rad}}{\text{ns}}$ and a time width of 141.41\,ns. The mean and standard deviation for (c) are 0.0338\,ns and 0.0533\,ns respectively. For (f) the mean is 0.1839\,kHz and the standard deviation is 0.9306\,kHz.}
    \label{ErrorHistograms_5MHzCase}
\end{figure*}
\begin{figure*}[ht!]
    \centering
    \includegraphics[width=\textwidth]{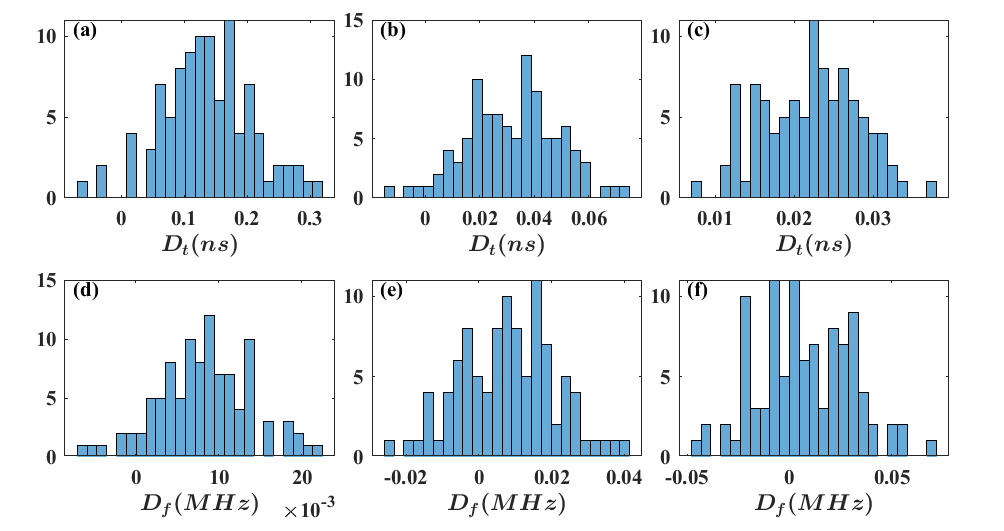}
    \caption{Time and frequency error analysis for the case where the input pulse has a frequency bandwidth of 100 MHz. Each histogram has 25 bins and 100 data points. The parameters for the plots are as follows: (a)(d) The pulses used have a chirp rate of $\Omega$'= 0.5 $\frac{\text{MHz}\,\text{Rad}}{\text{ns}}$ and a time width of $\delta_t$ = 533.63\,ns. The mean and standard deviation for (a) are 0.1360\,ns and 0.0713\,ns respectively. For (d) the mean is 0.0083\,MHz and the standard deviation is 0.0056 MHz. (b)(e) The pulses have a chirp rate of 5 $\frac{\text{MHz}\,\text{Rad}}{\text{ns}}$ and a time width of 53.23\,ns. The histogram in (b) has a mean of 0.0320\,ns and a standard deviation of 0.0170\,ns. The analogous histogram (e) has a mean of 0.0079\,Mhz and standard deviation of 0.013\,MHz. Lastly, for (c) and (f) the pulses used have a chirp rate of 28.48\,$\frac{\text{MHz}\,\text{Rad}}{\text{ns}}$ and a time width of 8.38\,ns. The mean and standard deviation for (c) are 0.0221\,ns and 0.0060\,ns respectively. For (f) the mean is 0.0072\,MHz and the standard deviation is 0.0233\,MHz.}
    \label{ErrorHistograms_100MHzCase}
\end{figure*}

To estimate the measurement statistical error in our experiments we measured a pulse sent through two identical cables. One cable connects Port 1+ on the AWG to channel 1 on the oscilloscope. The other cable connects Port 1- on the AWG to channel 2 on the oscilloscope. The output pulses measured from each cable are then compared by calculating the difference in center frequency and transmission time of the pulse. Ideally, these differences should be zero, however we see in Fig.\,\ref{ErrorHistograms_5MHzCase} and Fig.\,\ref{ErrorHistograms_100MHzCase} that that is not the case. We tested pulses with various parameters. In Fig.\,\ref{ErrorHistograms_5MHzCase} we use pulses that have a frequency bandwidth of 5\,MHz and chirp rates ranging from 0.008\,$\frac{\text{MHz}\,\text{Rad}}{\text{ns}}$ to 0.08\,$\frac{\text{MHz}\,\text{Rad}}{\text{ns}}$ with time widths ranging from 141.41\,ns to 1666\,ns. In Fig.\,\ref{ErrorHistograms_100MHzCase} we use pulses that have a frequency bandwidth of 100\,MHz and chirp rates ranging from 0.5\,$\frac{\text{MHz}\,\text{Rad}}{\text{ns}}$ to 28.48\,$\frac{\text{MHz}\,\text{Rad}}{\text{ns}}$ with time widths ranging from 8.28\,ns to 533.63\,ns. 

\begin{figure*}[ht!]
    \centering
    \includegraphics[width=\textwidth]{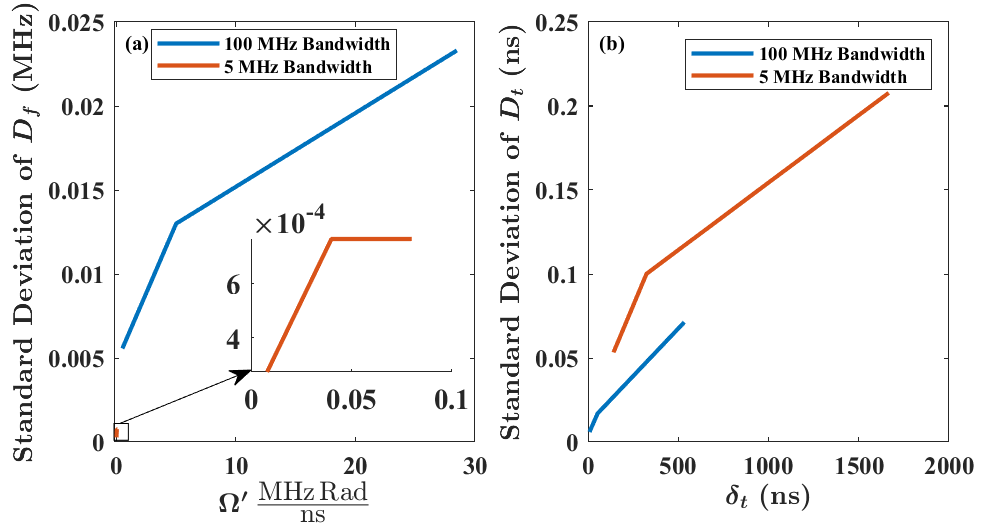}
    \caption{(a) Standard deviation of the time shift represented by the histograms in (a-c) of Figs.\,\ref{ErrorHistograms_5MHzCase} and \ref{ErrorHistograms_100MHzCase} versus the chirp rate of the input pulse used. The inset is the small boxed region in the lower left zoomed in. (b) Standard deviation of the frequency shift from the histograms in (d-f) of Figs.\,\ref{ErrorHistograms_5MHzCase} and \ref{ErrorHistograms_100MHzCase} versus the time width of the input pulse.} 
    \label{Error_SD_Plot}
\end{figure*}

For each case we repeated the measurement 100 times to create the distributions depicted in Figs.\,\ref{ErrorHistograms_5MHzCase} and \ref{ErrorHistograms_100MHzCase}. The standard deviations in the shift in frequency are plotted versus chirp rate in Fig.\,\ref{Error_SD_Plot}(a), and analogously the standard deviations in the shift in time versus the time width of the input pulse is shown in Fig.\,\ref{Error_SD_Plot}(b). The 5 MHz bandwidth case is plotted in orange and the 100 MHz pulse bandwidth case is plotted in blue. From these plots we clearly see that the standard deviation of $D_f$ increases with chirp rate and the standard deviation of $D_t$ increases with pulse width for both pulse frequency bandwidth cases. In Fig.\,\ref{Error_SD_Plot}(a) we see that the 100 MHz case consistently has a standard deviation that is larger than the smaller bandwidth case. Conversely, in Fig.\,\ref{Error_SD_Plot}(b) we see that the 5 MHz case consistently has a larger standard deviation than the 100 MHz case.

Since the standard deviations for the different bandwidth cases are different, we used different error bars depending on whether the pulse had a bandwidth that is closer to 5 MHz or 100 MHz. Note that the standard deviation in the shift in center frequency is so small that we decided not to include those error bars in our data plots. The error bars on the shift in time measurements are calculated by taking the average of the standard deviations of the corresponding histograms. For the $\sim$5 MHz cases we used a value of $\pm0.12$ ns for the error bars and for the $\sim$100 MHz cases we used a value of $\pm$0.03 ns. 
\clearpage

%\bibliography{References}

\end{document}